\documentclass[10pt]{article}
\usepackage{epsfig}
\usepackage{notconf}
\begin{document}

\setcounter{page}{127}

\title{Weighing 40 X-ray Luminous Clusters of Galaxies with Weak Gravitational Lensing}
\author{H. Dahle \\
    Institute for Astronomy, University of Hawaii}

\maketitle

\begin{abstract}
Deep two-color imaging of 40 clusters of galaxies with the NOT and the University of Hawaii 2.24-m telescope 
is used to measure the weak gravitational shear acting on faint background galaxies. 
From this, maps of the projected cluster mass distribution are constructed, and the cluster masses are measured within circular apertures of up to $\sim 3h^{-1} $~Mpc. The results are used to derive the cluster mass function at $z \sim 0.2$. 
The average mass-to-light ratio of the clusters indicate a low-density Universe with $\Omega_0 \simeq 0.25$.  
Additional mass concentrations are found along the lines of sight to the clusters, 
some of which are most likely associated with clusters at $z > 0.5$. 
In addition, the data are used to measure weak gravitational lensing due to large-scale structures 
along the line of sight. 
The future prospects for this kind of research with the NOT are briefly discussed.
  
\end{abstract}

\section{Introduction}

The weak gravitational lensing phenomenon has for a decade been used to probe the dark matter 
distribution in the Universe on scales ranging from galaxies to large-scale structures.
Rapid progress has been made both observationally 
and theoretically, and many important developments have taken place only 
during the last 2-3 years. Much of the progress on the observational side is due to the 
development of large-format (usually mosaic) CCD cameras which can cover fields of order 0.5 degrees 
on a side on 4-meter class telescopes. 
Such cameras are currently among the most heavily used instruments at leading observatories, 
and significant resources are being spent on building even larger CCD mosaics. 
Weak lensing studies are frequently mentioned as an important part of the scientific motivation  
for these technical developments.  

On the theoretical side, many different inversion algorithms for reconstructing the mass distribution from 
measurements of weak gravitational shear have been developed, e.g. to take into account strong lensing in the inner regions of the clusters.
With the increase in the size of CCD cameras, weaker and weaker shear can in principle be measured, provided that systematic effects can be kept under control. Recently developed algorithms can account 
for the effects of realistic PSFs and optimally weigh background galaxies of different fluxes, shapes and 
sizes when estimating the weak shear. 

Why all the interest in weak lensing? For the first time, we are getting an 
unbiased (although still fairly noisy) look at the mass distribution in the Universe. 
We are now at a stage where measurements are getting sufficiently sensitive to map  
the projected mass distribution along any line of sight. 
In many ways, this is similar to what happened in the past when new parts of the electromagnetic 
spectrum became available for astrophysical studies, but the mass probed by lensing is a 
much more fundamental quantity than electromagnetic radiation emission and this makes 
weak lensing measurements a very powerful tool for distinguishing between different models for structure 
formation. 

\section{The cluster sample}

Our observational efforts at the NOT have been focused on measuring weak gravitational shear  
in cluster fields. 
Although many weak lensing studies over the past decade have targeted clusters of galaxies -- see e.g.~the recent reviews by Mellier (1999) and Bartelmann \& Schneider (2000) and references therein -- 
they have mostly been limited to 
studies of a few clusters which are thought to be extremely massive, thereby guaranteeing 
a significant weak lensing signal. Until now, there has been no weak lensing study of a large, 
well-defined sample of objects. The sample described here was selected from the lists 
of X-ray luminous clusters published by Briel \& Henry (1993), Ebeling et al. (1996; XBACS) and 
Ebeling et al. (1998; BCS). The clusters are in the redshift range $0.15 < z < 0.35$ and 
almost all have X-ray luminosities L$_{\rm X, 0.1-2.4 keV} \geq 10^{45}$ erg s$^{-1}$ (for $h=0.5$, $q_{0}=0.5$). 
Most of the clusters are targets for recent or planned X-ray observations 
with Chandra or XMM-Newton. 

\begin{figure}
\centerline{\epsfig{file=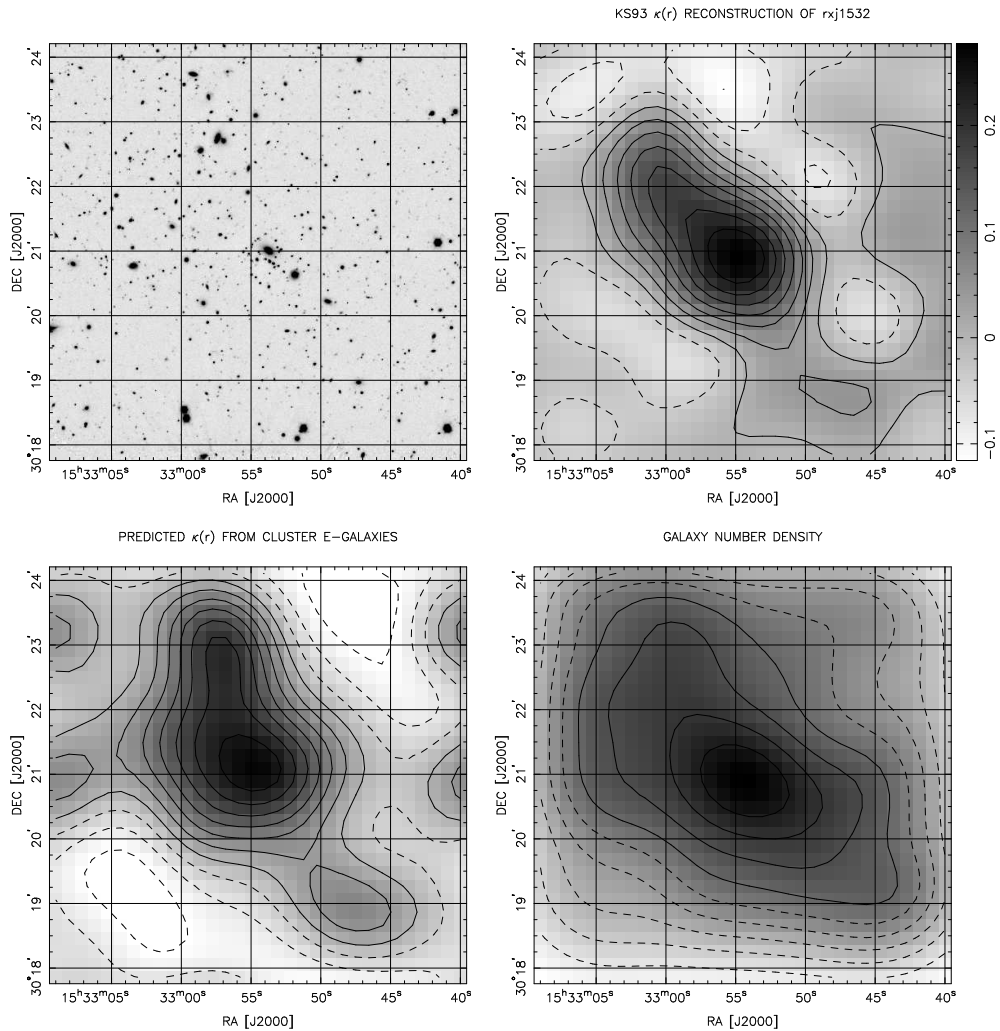}}
\caption{The upper left panel shows an I-band image of the cluster RXJ1532.9+3021 (z=0.345), based on 3 hours of exposure with ALFOSC in May 1999. The upper right panel shows the projected mass density in the field, calculated from weak lensing measurements. The lower left panel shows the predicted mass density, based on photometry of early-type cluster galaxies, and the lower right panel shows the number density of galaxies in the field. The contour plots have all been smoothed with a Gaussian of scale 36", and the average value has been subtracted from the two lower plots.}  
\label{fig1}
\end{figure}

\section{Observations and reduction}

The NOT observations were made with ALFOSC in imaging mode, and 1.5h of 
exposure was typically obtained for each cluster in each of the V- and I-bands. 
Additional observations were made with the University of Hawaii 2.24-m telescope 
at Mauna Kea Observatory with either a 2k Tektronix CCD camera 
or the UH8K CCD mosaic (Luppino et al. 1995). The UH8K pixels were $2\times 2$ re-binned
to effectively make a 4k detector with a 19' field.  To avoid introducing artificial shape distortions 
mimicking the effect of weak lensing, the individual exposures had to be accurately 
aligned to produce a deep combined exposure of each field. For the 2k data, catalogs of stellar 
positions was generated for all exposures, and an exposure with minimal extinction was chosen to be the 
astrometric and photometric reference for the other exposures. A least-squares solution was then calculated 
for a lower-order polynomial astrometric transformation of each exposure into the reference, and 
the images were median combined into a final image. To eliminate artificial galaxy shape distortions due to 
PSF anisotropies, a spatially variable PSF model was calculated, and the image was convolved with 
the model, rotated by $\pi/2$, to yield a smoothed image with a nearly circular PSF. 
The PSF model, along with the combined image and the smoothed image, was used to estimate the gravitational 
shear acting on a sample of faint background galaxies in the field with optimized weighting of the galaxies 
as a function of their fluxes, sizes and shapes (for details on the shear estimator, see Kaiser 2000).   

The 8k data were reduced using a proceduree very similar to the one described by Kaiser et al. (1999). 
An astrometric solution was found which accounts for the CCD chip layout in the camera and field distortions 
(which were quite small), using the USNO-A2.0 catalog as a reference. The PSF is 
known to vary discontinuously across chip boundaries in the UH8K camera, so the individual chip frames were convolved with 
a re-circularizing filter generated as described above before combining the frames. A set of `raw' and 
`re-circularized' combined images were then made, and the shear was estimated in essentially the same 
way as for the 2k data. 

\section{Weak Lensing Analysis} \label{WLanal}
 
\subsection{Predicted and reconstructed surface density}

Using local measurements of the spectral energy distributions (SEDs) of galaxies of different types 
(Coleman et al. 1980), and assuming negligible galaxy color evolution, it is possible to predict the 
broadband colors of galaxies as a function of redshift.  
Lubin (1996) shows that the colors of early-type galaxies at redshifts $z < 0.4$ are well represented 
by this no-evolution prediction. In this redshift range, early-type galaxies can be separated by their 
V-I colors from spirals and irregulars, and early-type galaxies in rich clusters form 
a tight, well-defined sequence in color-magnitude space. It is thus fairly easy to identify and isolate the early-type 
cluster galaxies and other early-type galaxies at moderate redshifts.  Assuming that the mass traces the 
early-type galaxies with a constant mass-to-(early type galaxy) light ratio, a map of the 
predicted mass distribution in the field can be generated and compared to the mass distribution 
derived from weak lensing. The projected surface density was reconstructed from the shear measurements 
using the algorithm of Kaiser \& Squires (1993), which is fast to implement and has well-understood noise 
properties. An example of the results for the cluster RXJ1532.9+3021 is shown in figure~\ref{fig1}.      

\subsection{Light-mass cross-correlation}
To investigate the relative distribution of mass and galaxy light in the field, the two-dimensional cross-correlation between the measured and predicted mass distribution was calculated. This was compared to the auto-correlation of the predicted mass distribution from galaxy photometry. The results of this procedure when applied to wide-field data is particularly interesting, since they probe a wide range of physical radii in the clusters. The clusters display a range of properties: In some clusters, the dark matter appears to be significantly more concentrated than the early-type galaxy light, in others the two components trace each other within the errors, and in some clusters the mass is significantly more extended than the light. The differences appear to be related to the dynamical state of the clusters. The clusters which appear to be completely relaxed systems have the more concentrated dark matter distributions, whereas the clusters with the more extended dark halos appear to still be in the later stages of their formation process. 

\begin{figure}
\centerline{\epsfig{file=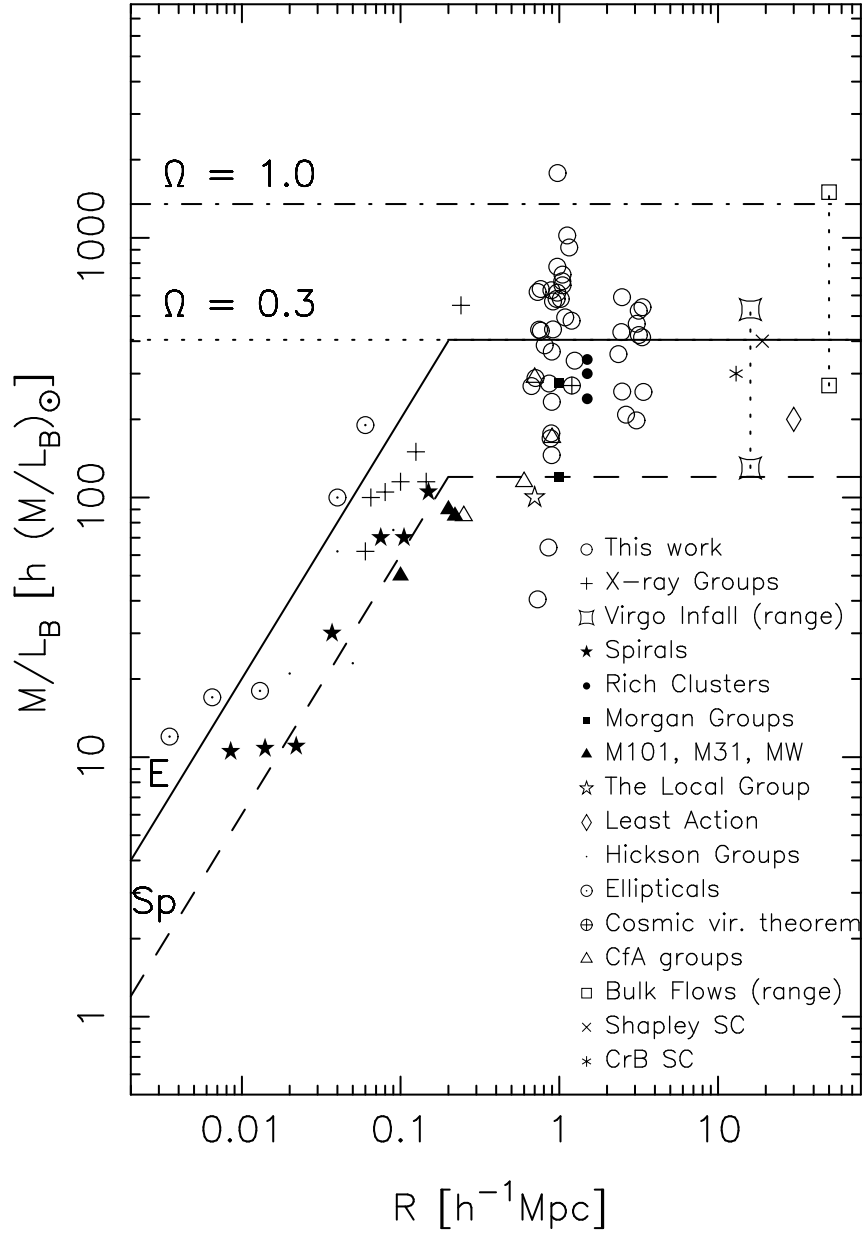}}
\caption{(Rest-frame) B-band mass-to-light ratio vs. physical scale.
The $M/L_B$ expected for $\Omega = 1$ and $\Omega = 0.3$ 
are shown as lines, along with empirical fits to $M/L_B$ (assumed to increase proportional 
to R up to a scale of $100 h^{-1}$ kpc) for spirals and ellipticals. The plotted points for 
rich clusters, Morgan groups, Hickson groups, CfA groups, spirals and ellipticals
 are median values of these samples. The circles are values for the clusters studied in this 
work, using the mass to total light ratio. The figure is in part based on a similar plot by 
Bahcall et al. (1995). See this paper for further references.}
\label{fig2}
\end{figure}

The rest-frame B-band mass-to-light ratios in the cluster fields were calculated by taking the discrete Fourier transform of the 
predicted and measured surface density images and comparing the amplitudes of the Fourier modes. In figure~\ref{fig2}, the resulting 
values for $M/L_{\rm B}$ are plotted against the physical scale corresponding to the longest wavelength measured. Also plotted are a 
number of other $M/L_{\rm B}$ measurements taken from the literature (Bahcall et al. 1995). The average value for the observed clusters 
is $M/L_{B} = 377 \pm 17$ in solar units, which coupled with local measurements of the B-band luminosity density (Loveday et al. 1992) 
indicates a value for the density parameter of $\Omega_0 \simeq 0.25$. 

\subsection{Comparison with other mass measurements}

\begin{figure}
\centerline{\epsfig{file=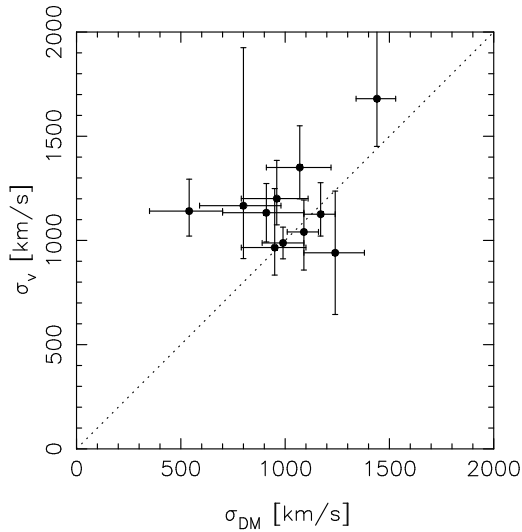,angle=-90}}
\caption{Spectroscopically measured velocity dispersion $\sigma_v$ vs. the dark matter velocity dispersion $\sigma_{\rm DM}$, inferred from weak lensing measurements.}
\label{fig3}
\end{figure}

Many of the clusters in the sample studied here have X-ray temperature measurements and/or galaxy velocity dispersion measurements available.
By fitting a singular isothermal sphere model to the measured tangential shear profiles around the clusters, the  
weak lensing measurements can be compared to velocity dispersion measurements, as shown in figure~\ref{fig3}. 
The agreement is generally fairly good, with a tendency for the spectroscopically measured velocity dispersion 
to be marginally larger than the dark matter velocity dispersion (derived from the weak lensing measurements) 
for about half of the clusters. This is not unexpected, since the spectroscopically measured velocity dispersions  
may be biased upwards due to line-of-sight contamination by galaxies in the infall region outside the 
virial radius of the clusters. 

\subsection{The mass function of galaxy clusters}

\begin{figure}
\centerline{\epsfig{file=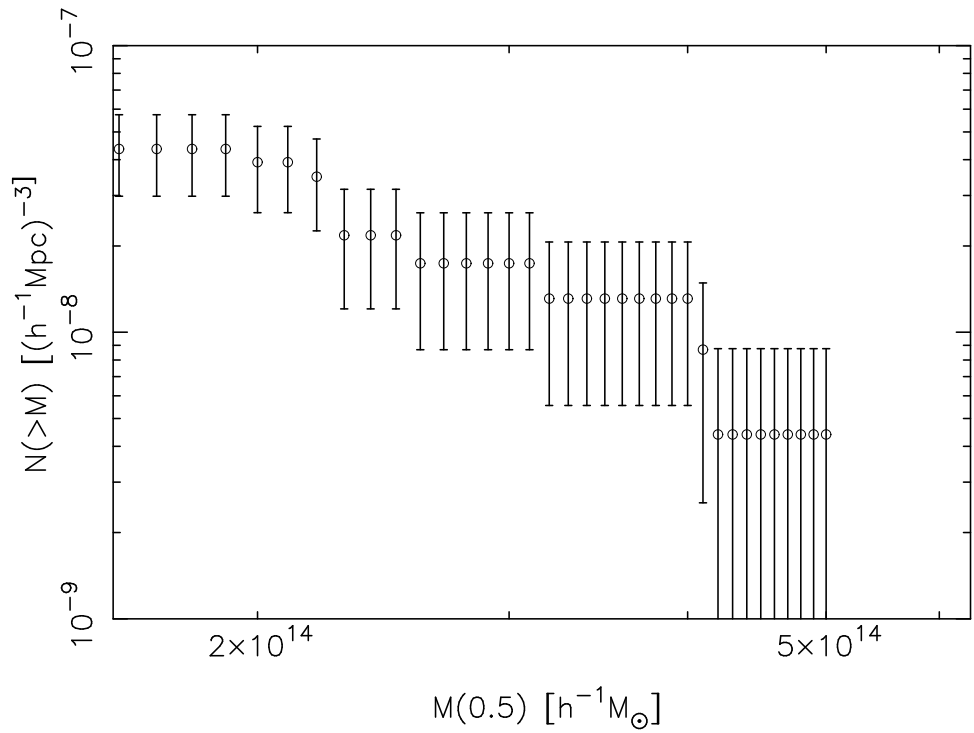,angle=-90}}
\caption{Cluster mass function. The quantity $M(0.5)$ is the mass within a cluster radius of $0.5h^{-1}$~Mpc in three-dimensional space.}
\label{fig4}
\end{figure}

From the list of observed clusters, a volume-limited BCS subsample with the selection criteria 
$0.1 < z < 0.25$ and L$_{\rm X, 0.1-2.4 keV} \geq 10^{45}$ erg s$^{-1}$ was selected. 
Using the singular isothermal sphere model fit, the masses contained 
within the inner $0.5 h^{-1}$~Mpc of the clusters were calculated, and a cumulative cluster mass function was constructed, 
as shown in figure~\ref{fig4}. This is the first mass function of galaxy clusters which is based on gravitational lensing measurements. 
Further work is currently in progress to compare the results to predictions from Press-Schechter theory and from recent large cosmological N-body simulations (e.g. Jenkins et al. 2000).  
 
\subsection{Detection of additional mass concentrations along the line of sight} 

\begin{figure}
\centerline{\epsfig{file=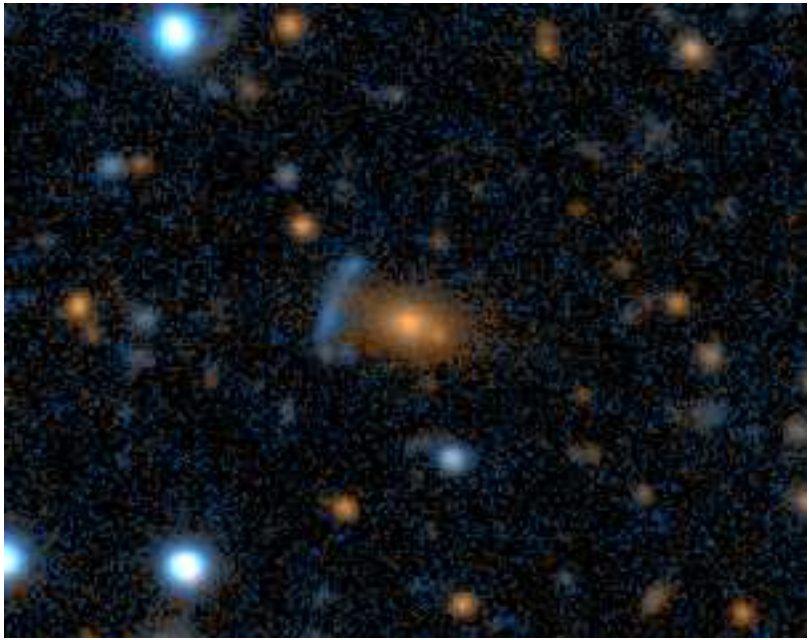}}
\caption{A blue arc curving around the central galaxy of a previously unknown, distant cluster.}
\label{fig5}
\end{figure}

In addition to the targeted clusters, some additional mass concentrations are seen in the reconstructed density maps of 
some cluster fields. These mass concentrations tend to be associated with concentrations of red ($V-I > 2.2$), faint galaxies, which  
indicates that they are distant clusters ($z > 0.5$). One distant cluster candidate, situated in the field of Abell 1705, even shows a strongly lensed blue arc surrounding what appears to be the central cluster galaxy (see figure~\ref{fig5}). Thus, the weak lensing measurements can be used to discover new clusters in the observed fields. The availability of color information is a great help in this respect: What initially appeared as a `dark cluster' in the field of Abell 1722, similar to the one reported by Erben et al. (2000), turned out to be associated with a concentration of red galaxies which did not stand out from the background/foreground in data from a single passband only.     

\subsection{Cosmic shear in cluster fields}

Several groups have recently reported the detection of weak cosmic shear from large-scale structures along the line of sight
(Bacon et al. 2000; Kaiser, Wilson \& Luppino 2000; van Waerbeke et al. 2000, Wittman et al. 2000). These studies are generally 
based on observations of `blank fields' not containing any previously known clusters. The presence of the massive clusters in the imaging data
described here may bias the cosmic shear measurement somewhat, but the shear on different sides of the cluster (which is generally 
situated near the center of each field) will mostly cancel out, and measurements of the net shear across the image will thus only 
be weakly affected by the presence of the cluster. Using 12 cluster fields observed in the I-band with the UH8K camera, the 
cosmic shear variance on scales of 20' was measured to be $\langle \bar{\gamma}^2 \rangle_{II} = 1.24 \pm 0.49 \times 10^{-4}$, 
consistent with the results of Wittman et al., but marginally larger than the Kaiser et al. result. Using a cross-correlation of shear measurements from I-band data and shallower V-band data in 7 UH8K fields (which will cancel out systematic effects which are not common to both passbands), an upper limit of $\langle \bar{\gamma}^2 \rangle_{IV} < 0.50 \times 10^{-4}$ was derived, more in line with the values found by Kaiser et al. 

\section{Future prospects: What NOT can (and can't) do}

Although the Mosaic/FRED combination would still be extremely competitive if both instruments were 
already available, the NOT will within a couple of years become less attractive for weak lensing studies 
requiring very wide fields (e.g. measurements of cosmic shear and weak galaxy-galaxy lensing), since 
the field sizes of competing facilities (such as Subaru Suprime-Cam, CFHT MegaPrime, 
MMT Megacam, VST and VISTA) are rapidly increasing. In the more distant future, wide-field weak lensing observations
will most likely be carried out using dedicated facilities 
such as the proposed Dark Matter Telescope (see {\tt http://www.dmtelescope.org}) or an array of smaller 
telescopes (Kaiser, Tonry \& Luppino 2000). Until then, the NOT (with Mosaic/FRED) will likely remain competitive for weak lensing 
observations that target a fairly large number of objects along widely separated lines of sight, such as studies of known galaxy 
clusters and superclusters. The main advantages of the NOT will still be the excellent intrinsic seeing at 
the telescope site (work to further improve image quality should have very high priority !!) and the possibility for 
relatively extensive telescope time allocations. 

\acknowledgments

The author wishes to thank the organizers of this meeting for travel support, and 
he also wishes to thank Gillian Wilson for many useful discussions.
The work presented here is part of a larger effort to study galaxy clusters, in collaboration with 
Ragnvald Irgens, Per B. Lilje, Nick Kaiser, Steve Maddox and Kristian Pedersen.  
The author has been sponsored by a Ph.D. research stipend awarded by 
Norges Forskningsr{\aa}d (project number 110792/431).

\end{document}